\documentclass[prl, superscriptaddress, twocolumn, showpacs]{revtex4}

\usepackage{amsfonts}
\usepackage{amssymb}
\usepackage[dvips]{graphicx}
\usepackage{amsmath}

\newcommand{\bra}[1]    {\langle #1|}
\newcommand{\ket}[1]    {| #1 \rangle}
\newcommand{\e}         {\mathrm{e}}
\newcommand{\ii}         {\mathrm{i}}

\begin{document}

\title{Optimal Quantum Phase Estimation}

\author{U.~Dorner}
\affiliation{Clarendon Laboratory, University of Oxford, Parks Road, Oxford OX1 3PU, United Kingdom}
\author{R.~Demkowicz-Dobrzanski} 
\affiliation{Institute of Physics, Nicolaus Copernicus University, Grudziadzka 5, PL-87-100 Toru\'{n}, Poland}
\author{B.~J.~Smith} \author{J.~S.~Lundeen} 
\affiliation{Clarendon Laboratory, University of Oxford, Parks Road, Oxford OX1 3PU, United Kingdom}
\author{W.~Wasilewski} 
\affiliation{Institute of Experimental Physics, University of Warsaw, Ho\.{z}a 69, PL-00-681 Warsaw, Poland}
\author{K.~Banaszek}
\affiliation{Institute of Physics, Nicolaus Copernicus University, Grudziadzka 5, PL-87-100 Toru\'{n}, Poland}
\author{I.~A.~Walmsley}
\affiliation{Clarendon Laboratory, University of Oxford, Parks Road, Oxford OX1 3PU, United Kingdom}

\date{\today}

\begin{abstract} 
  By using a systematic optimization approach we determine quantum
  states of light with definite photon number leading to the best
  possible precision in optical two mode interferometry. Our treatment
  takes into account the experimentally relevant situation of photon
  losses. Our results thus reveal the benchmark for precision in
  optical interferometry. Although this boundary is generally worse
  than the Heisenberg limit, we show that the obtained precision beats
  the standard quantum limit thus leading to a significant improvement
  compared to classical interferometers. We furthermore discuss
  alternative states and strategies to the optimized states which are
  easier to generate at the cost of only slightly lower precision.
\end{abstract}

\maketitle

Interferometry is one of the most important measurement techniques in
physics. Its numerous variations include Ramsey spectroscopy in atomic
physics, optical interferometry in gravitational wave detectors, laser
gyroscopes or optical imaging to name but a few. All these
applications aim to estimate the quantity of interest, normally a
relative phase gathered by one ``arm'' of the interferometer, with
highest possible precision.  In this letter we present fundamental
limits to this precision in optical interferometry for light with
definite photon number in the presence of losses.  To find these
limits it is necessary to consider the ``cost'' of the experiment,
i.e. the required resources, and determine the precision of the
estimated phase as a function of the cost.  In optical interferometry
the required resource is typically identified to be the number of
photons $N$ necessary to reach a desired precision.  Classically the
precision of the estimated phase scales then like $1/\sqrt{N}$, the so
called standard quantum limit (SQL).  Quantum interferometry on the
other hand promises to beat this limit by employing highly
non-classical entangled states to drastically improve the precision to
a scaling $1/N$ known as the Heisenberg
limit~\cite{Giovanneti04,Giovanneti06}. The realization of
interferometric measurements beyond the SQL is a very active field and
recent years have seen tremendous
progress~\cite{Mitchell04,Eisenberg05,Nagata07,Resch07,Higgins07}.  A
quantum enhancement in precision would allow for a significant
reduction of the energy flux while keeping the same measurement
precision. This is important, for example, if the phase is induced by
a fragile sample~\cite{Higgins07}. However, most of the theoretical
work done so far ignores the unavoidable presence of noise in the
system. Existing treatments come to the conclusion that the benefit
from highly entangled states deteriorates quickly even if only a small
amount of noise is present in the
system~\cite{Huelga97,Shaji07,Sarovar06,Gilbert06,Huver08}. This is
not really a surprise since states of this type are typically very
fragile: In optical interferometry, the well-studied N00N state
promises to provide Heisenberg limited sensitivity in phase
estimation~\cite{Bollinger96}, however, the loss of merely a single
photon renders this state useless since it collapses into a product of
two Fock states which cannot acquire any phase information.

The Heisenberg limit is believed to be the ultimate precision in
optical phase estimation, however, it is yet an unsolved problem if
this limit can be reached in the presence of noise and, if not, then
what is the ultimate precision? In this letter we answer this question
for optical two-mode interferometry in the presence of photon losses,
which is the limiting source of noise in such experiments.  By using a
systematic approach we determine optimal states with definite photon
number leading to the highest possible precision. Although it turns
out that the Heisenberg limit is unattainable we show that one can
beat the SQL thus greatly improving precision beyond classical
interferometry. Furthermore we introduce alternatives to the optimal
states, with simpler structure, at the cost of only slightly less
precision.
\begin{figure}[ht]
\centering\includegraphics[width=6.1cm]{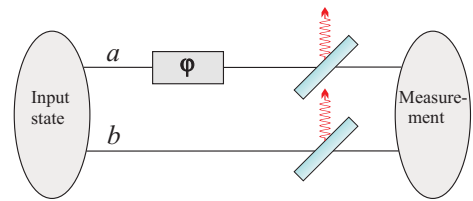}
\caption{%
  General optical interferometric setup with two arms. Channel $a$
  acquires a phase $\varphi$ relative to channel $b$. The two beam
  splitters in channel $a$ and $b$ symbolize photon losses.}
\label{fig1}
\end{figure}

We consider a general interferometer with two arms as shown in
Fig.~\ref{fig1}, in particular we do not put any restrictions on the
measurement scheme. Channel $a$ is accumulating a phase $\varphi$
relative to channel $b$ and both arms, $a$ and $b$, are subject to
photon loss which can be seen as the effect of fictitious beam
splitters inserted at arbitrary locations in both channels. We aim to
estimate $\varphi$ with the highest possible precision quantified by
using the uncertainty of the estimated phase $\varphi_{est}$,
\begin{equation}
\delta\varphi = \left\langle \left(\varphi_{est}{\left|{\partial\langle \varphi_{est}\rangle}/{\partial\varphi}\right|^{-1}} - \varphi\right)^2  \right\rangle^{\frac{1}{2}},
\end{equation}
which, for an unbiased estimator, is simply the standard deviation.
According to the general theory of quantum parameter
estimation~\cite{Braunstein94,Helstrom,Holevo} $\delta\varphi$ is
bounded by the quantum Cram\'er-Rao bound
\begin{equation}
\delta\varphi \ge \frac{1}{\sqrt{\nu}\sqrt{F_Q}}\equiv\frac{\delta\varphi_{min}}{\sqrt\nu},
\label{eq:error}
\end{equation}
where $\nu$ is the number of experimental runs and $F_Q$ is the
quantum Fisher information. It was shown that this bound can always be
reached asymptotically by maximum likelihood estimation and a
projective measurement in the eigenbasis of the `symmetric logarithmic
derivative operator'~\cite{Braunstein94}. Hence,
inequality~(\ref{eq:error}) defines the {\em principally smallest
  possible uncertainty} in phase estimation the determination of which
is the primary scope of this paper. An explicit construction of the
measurement operators will be given elsewhere~\cite{Demkowicz08}.

Photon losses can be modelled by inserting fictitious beam splitters
with transmissivities $\eta_{a,b}$ into both arms of the
interferometer which couple the system to (uncorrelated) environments.
The noise operation and the phase accumulation commute, i.e. it is
irrelevant if photons are lost before, during or after channel $a$
acquires its relative phase with respect to $b$. If the noise
operation is applied first, the state after tracing out the
environmental degrees of freedom can be written as $\rho =
\sum_{k,l=0}^\infty K_{l,a} K_{k,b}\rho_{in} K_{k,b}^\dagger
K_{l,a}^\dagger$ with Kraus operators $K_{l,a} =
(1-\eta_a)^\frac{l}{2}\eta_a^{\frac{1}{2}\hat a^\dagger \hat a} \hat
a^l/\sqrt{l!}$, where $\hat a$ is the annihilation operator for mode
$a$, and analogously for mode $b$. This state acquires a phase through
the transformation $\rho(\varphi) = \e^{-\ii\varphi \hat a^\dagger
  \hat a} \rho \e^{\ii\varphi \hat a^\dagger \hat a}$. This scenario
is equivalent to a continuous photon loss model described by a master
equation with loss rates $|\ln\eta_{a,b}|$ per unit time.

We consider the most general pure input states with definite photon
number $N$,
\begin{equation}
\ket{\psi}_{in} = \sum_{k=0}^N \alpha_k \ket{k,N-k},
\label{eq:input}
\end{equation}
where $\ket{k,N-k}$ abbreviates the Fock state $\ket{k}_a\ket{N-k}_b$.
Special cases of (\ref{eq:input}) comprise in particular the highly
entangled N00N state, $(\ket{N0}+\ket{0N})/\sqrt2$ which, in the
absence of noise, leads to Heisenberg limited phase estimation but is
strongly prone to decoherence otherwise. Equation~(\ref{eq:input})
includes states which are ``less'' entangled but more robust
representing a trade-off between precision and robustness. Also, in
the absence of additional reference beams, a superposition of states
with different definite photon number would effectively become a
mixture~\cite{Molmer97} of these states. Convexity of
$F_Q$~\cite{Fujiwara01} implies then that the analysis can be
restricted to states with definite photon number if we use them
successively~\cite{Demkowicz08}.

In the case of no losses, the state of the system,
$\ket{\psi(\varphi)}=\e^{-\ii\varphi \hat a^\dagger \hat
  a}\ket{\psi}_{in}$, remains pure and the quantum Fisher information
reads
\begin{equation}
\label{eq:FQpure}
F_Q = 4[\langle\psi^\prime(\varphi)|\psi^\prime(\varphi)\rangle - |\langle\psi^\prime(\varphi)|\psi(\varphi)\rangle|^2]
=4(\Delta(\hat a^\dagger \hat a))^2, 
\end{equation}
where $(\Delta(\hat a^\dagger \hat a))^2$ is the variance of $\hat
a^\dagger \hat a$ with respect to the state $\ket{\psi}_{in}$ and the
prime indicates a derivative with respect to
$\varphi$~\cite{Braunstein94}.  In the presence of noise the pure
input state will deteriorate into a mixture $\rho(\varphi)$. If the
eigenvalues and eigenstates of $\rho(\varphi)$ are known the quantum
Fisher information can be easily calculated~\cite{Braunstein94}.
However, very often the analytical diagonalization of $\rho(\varphi)$
turns out not to be feasible. In this case, if the density operator is
given in the form $\rho(\varphi)=\sum
p_j\ket{\psi_j(\varphi)}\bra{\psi_j(\varphi)}$, where the
$\ket{\psi_j(\varphi)}$ are not necessarily orthogonal, we can use the
convexity of $F_Q$~\cite{Fujiwara01} to obtain an upper bound
\begin{equation}
F_Q\le \tilde F_Q =  4\sum_j p_j (\Delta(\hat a^\dagger \hat a)_j)^2,
\label{eq:FQ_ineq}
\end{equation}
where the variance corresponds to $\ket{\psi_j(\varphi)}$.  The bound
is reached if the spaces spanned by
$\{|\psi_j(\varphi)\rangle,|\psi'_j(\varphi)\rangle\}$ and
$\{|\psi_i(\varphi)\rangle,|\psi'_i(\varphi)\rangle\}$ are orthogonal
for $j \neq i$. Particularly, we have $F_Q=\tilde F_Q$ for the $N00N$
state and, generally, if photon losses are only present in one
channel, i.e., $\eta_b=1$.  The latter is relevant if the phase
$\varphi$ is induced by a sample in arm $a$ which also causes the
majority of photon losses. Applying Eq.~(\ref{eq:FQ_ineq}) to the
state~(\ref{eq:input}), we get
\begin{equation} 
\tilde F_Q = 4\left( 
\sum_{k=0}^N k^2 x_k - \sum_{l=0}^N\sum_{m=0}^{N-l}\frac{\left( \sum_{k=l}^{N-m} x_k k B^k_{lm} \right)^2}{\sum_{k=l}^{N-m} x_k B^k_{lm}}
\right)
\label{eq:fisherbound}
\end{equation}
with $x_k=|\alpha_k|^2$ and $B^k_{lm} \equiv
\binom{k}{l}\binom{N-k}{m}\eta_a^k(\eta_a^{-1}-1)^{l}\eta_b^{N-k}(\eta_b^{-1}-1)^{m}$.
For $\eta_b=1$ we have
\begin{equation} 
\tilde F_Q = F_Q = 4\left( 
\sum_{k=0}^N k^2 x_k - \sum_{l=0}^N\frac{\left( \sum_{k=l}^{N} x_k k B^k_{l} \right)^2}{\sum_{k=l}^{N} x_k B^k_{l}}
\right)
\label{eq:fisherbound2}
\end{equation}
with $B^k_{l}\equiv B^k_{l0}$. Obviously, the phases of the $\alpha_k$
are irrelevant. Furthermore, we proved analytically that $\tilde F_Q$
and $F_Q$ are concave functions of the $\{x_k\}$~\cite{Demkowicz08}.
This simplifies the numerical maximization of $\tilde F_Q$ or $F_Q$
and more importantly, it implies that any maximum is global.

\begin{figure}[t]
\centering\includegraphics[]{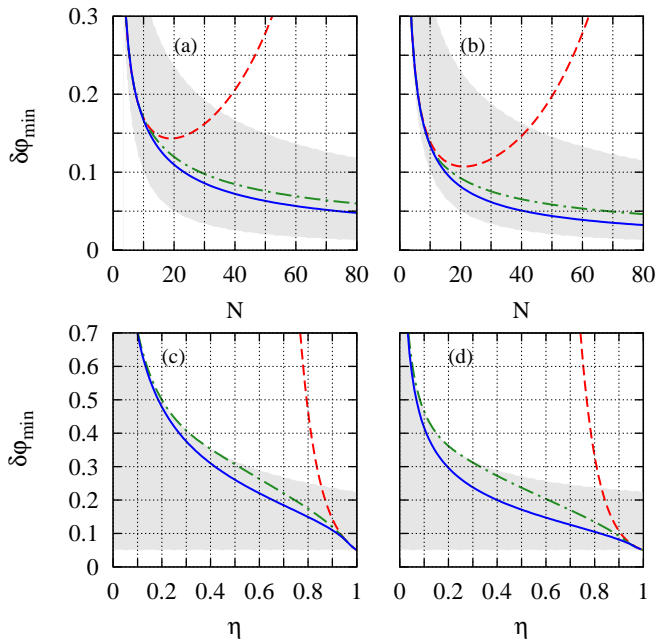}
\caption{%
  Phase estimation precision $\delta\varphi_{min}$ for
  losses in both arms of the interferometer (a) versus photon number
  $N$ ($\eta_a=\eta_b=\eta=0.9$) and (c) versus transmissivity
  $\eta_a=\eta_b=\eta$ ($N=20$). The precision for losses in one arm,
  i.e. $\eta_b=1$, is shown in (b) versus $N$ ($\eta_a=\eta=0.9$) and
  in (d) versus $\eta_a=\eta$ ($N=20$). Blue, solid line: Optimal state; Red,
  dashed line: N00N state; Green, dashed-dotted line: N00N chopping strategy; Shaded area:
  Region between Heisenberg limit and classical limit (see text).  }
\label{fig2}
\end{figure}

Figure~\ref{fig2} shows the results of such an optimization for
$\eta_a=\eta_b\equiv\eta=0.9$ and $\eta_a\equiv\eta=0.9,\,\eta_b=1$,
i.e. $10\%$ losses in both arms and one arm, respectively (blue, solid
lines). In the following we concentrate on these two scenarios.  The
quantity we analyze is $\delta\varphi_{min}\equiv 1/\sqrt{F_Q}$ (or
$1/\sqrt{\tilde F_Q}$) corresponding to the best measurement precision
for fixed $\nu$ [see Eq.~(\ref{eq:error})]. The lower and upper
boundaries of the shaded regions in Fig.~\ref{fig2} are the Heisenberg
limit, $1/N$, and a {\em standard interferometric limit} (SIL)
\footnote{The SIL corresponds to a Mach-Zehnder interferometer with a
  coherent state of unknown phase and the vacuum at the input ports.
  This state is equivalent to a statistical mixture of Fock states
  with Poissonian distribution. For losses in one arm the beam
  splitter has to be unbalanced to improve the precision. We note that
  the SIL is also obtainable by sending successively $N$ single
  photons through the interferometer.} given by $1/\sqrt{N\eta}$
(losses in both arms) and $(1+\sqrt{\eta})/2\sqrt{N\eta}$ (losses in
one arm). Since the SIL is obtained by a classical reference
experiment (it scales like the SQL), $\delta\varphi_{min}$ falling
into the shaded region implies an improvement over a classical
interferometer. For $\eta_b\ne 1$ we used the state which maximizes
$\tilde F_Q$ to calculate $F_Q$ which differed by no more than
$0.45\%$. Due to Eq.~(\ref{eq:FQ_ineq}) the ``true'' maximum has to
lie in between these quantities and its deviation from $\tilde F_Q$
can be neglected on the scale given by the difference of the SIL and
the Heisenberg limit.
As can be seen in Figs.~\ref{fig2}(a)(b) it is obviously not possible
to reach the Heisenberg limit using input states with definite photon
number.  However, we gain a significant improvement over the SIL of up
to $60\%$ (losses in both arms) and $73\%$ (losses in one arm). The
precision for the N00N state in the presence of losses [dashed lines
in Figs.~\ref{fig2}(a)(b)] is given by
$\delta\varphi_{min}=1/N\eta^{N/2}$ for losses in both arms and ceases
to be optimal for $N>7$ photons.  For losses in one arm we have
$\delta\varphi_{min}=\sqrt{1+\eta^{-N}}/\sqrt 2N$ which is generally
worse than the optimal state: Here it is beneficial to use N00N states
with unequal amplitudes of the two components. The best precision of
such an ``unbalanced'' N00N state is given by $(1+\eta^{-N/2})/2N$
which coincides with the optimal state for $N<10$. However, for larger
photon numbers N00N states are not preferable; the precision gets even
worse than the SIL.

\begin{figure}[t]
\centering\includegraphics[]{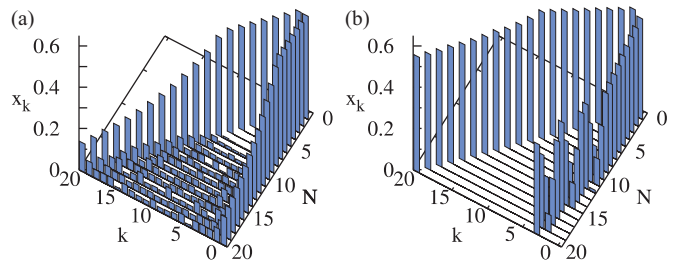}
\caption{%
  Coefficients $x_k=|\alpha_k|^2$ of the optimal state
  versus photon number $N$. (a) losses in both arms
  ($\eta_a=\eta_b=\eta=0.9$). (b) losses in one arm
  ($\eta_a=\eta=0.9,\,\eta_b=1$).}
\label{fig2a}
\end{figure}

Figure~\ref{fig2a} shows that the optimal state for losses in both
arms has generally many non-zero components. Intuitively this is
consistent with the idea that the loss of a photon does not radically
change the photon number distribution. The structure of the optimal
state is simpler for losses in one arm. We therefore compare it to
states with only two non-zero components.  The best precision obtained
by these states differs by no more than $3\%$ from the optimal case
for the example shown in Fig.~\ref{fig2}(b). They have the form
$\sqrt{p}\ket{m,N-m} + \sqrt{1-p}\ket{N,0}$ and are thus similar to
the optimal state. This reflects the fact that it is both beneficial
to have a large photon number difference between arm $a$ and $b$ and
have $m>0$ so that loss of a photon does not completely destroy the
coherence. For equal losses in both arms, the best two-component state
has approximately a symmetric form $(\ket{m,N-m} +
\ket{N-m,m})/\sqrt2$, but the corresponding precision deviates
significantly from the optimal state [up to $13\%$ for the example
shown in Fig.~\ref{fig2}(a)]. Here, states with more non-zero
components are more useful, e.g. a twin Fock state~\cite{Holland93}
reducing the difference to $9\%$.  Figure~\ref{fig2}(c) and~(d) shows
the best possible precision versus $\eta$ for $N=20$.  For
$\eta\gtrsim 0.95\approx\e^{-1/N}$ the optimal state, the optimal
two-component state and the (unbalanced) N00N state are identical.

We can also use different strategies to operate the interferometer:
Since our resources are given by the total number of photons, $N$, we
can, instead of employing a single N00N state, use these photons to
generate $N/n$ n00n states containing $n\le N$ photons each, i.e.
we split up a ``larger'' N00N state into a number of ``smaller'' n00n
states which are sent successively through the interferometer.
Maximization over $n$ (treated as a continuous parameter) leads to the
precision
\begin{equation}
\delta\varphi_{min} =
\begin{cases}
  \frac{1+\sqrt{\tilde\eta}}{2\sqrt{N\eta}} & ;\eta\le \eta_0^{-1}\\
  \frac{1+\sqrt{\tilde\eta_0}}{2\sqrt{N\tilde\eta_0}}\sqrt{\frac{\eta_0|\ln{\eta}|}{\ln{\eta_0}}} & ; \eta_0^{-1} <\eta\le \eta_0^{-\frac{1}{N}}\\
  \frac{1+\tilde\eta^{\frac{N}{2}}}{2 N\eta^{\frac{N}{2}}} & ;\eta >
  \eta_0^{-\frac{1}{N}},
\end{cases}
\label{eq:chop2}
\end{equation}
where $\tilde\eta=\tilde\eta_0=1,\,\eta_0=\e$ and
$\tilde\eta=\eta,\,\tilde\eta_0=\eta_0\approx 4.386$ for losses in
both arms and one arm, respectively. In the latter case we use
unbalanced n00n states. Examples for this ``chopping'' strategy are
given by the green, dashed-dotted lines in Fig.~\ref{fig2}. Note that in this case
the total number of experimental runs is $\nu N/n$.  The number of
photons ``per run'' is $n = 1$, $n = \ln{\eta_0}/|\ln{\eta}|$ and
$n=N$ for the three cases in Eq.~(\ref{eq:chop2}).  For sufficiently
small total photon numbers [cf. last line in Eq.~(\ref{eq:chop2})], it
is not an advantage to chop the N00N state. If $N$ is larger, the
strategy does not improve the scaling with $N$ compared to the SIL (or
SQL).  Nonetheless, it is an improvement over the SIL by a constant
factor of approximately 2 (losses in both arms) or 2.5 (losses in one
arm) in the example shown in Fig.~\ref{fig2}, i.e. we need almost four
(6.2) times less photons to reach the same measurement precision.
\begin{figure}[t]
  \centering\includegraphics[]{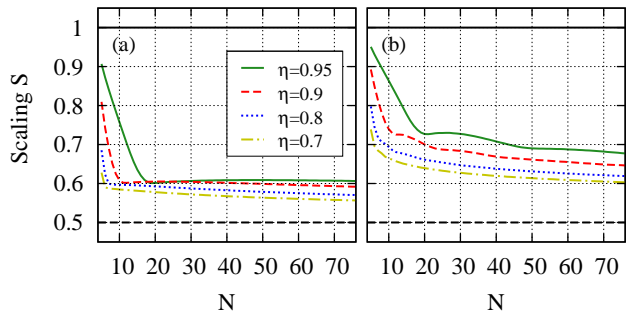}
\caption{%
  Differential scaling of phase estimation precision
  $\delta\varphi_{min}$ with number of photons $N$ for different
  transmissivities. (a) Optimal state for losses in both arms
  ($\eta_a=\eta_b\equiv\eta$) and (b) for losses in one arm
  ($\eta_a\equiv\eta,\,\eta_b=1$). The color coding is the same in
  both plots.}
\label{fig3}
\end{figure}

The scaling of the precision of the optimal state with the number of
photons turns out not to behave exactly like a power law. Therefore we
define a differential scaling $S(N)$ given by the local slope of
$\delta\varphi_{min}(N)$ on a log-log scale obtained by a linear fit
to the points corresponding to $N-4,\cdots,N+4 $. If
$\delta\varphi_{min}$ scales like a power law $S$ would be the
constant power, e.g. $S=0.5$ for the SQL and $S=1$ for the Heisenberg
limit. Results are shown in Fig.~\ref{fig3}.  It is clearly visible
that the scaling of the optimal state drops quickly, tending away from
the Heisenberg limit towards the SQL. Even for rather high
transmissivity (green lines correspond to $95\%$) a scaling of $\sim
0.61$ (losses in both arms) or $\sim 0.68$ (losses in one arm) for
$N=70$ is not exceeded.  Moreover, the scaling gets worse for higher
photon numbers. As yet it remains an unresolved but fascinating
question if the scaling of the optimal states eventually tend to the
SQL for all $\eta<1$ or asymptotically reaches a value which beats the
SQL proving a true quantum scaling advantage in the presence of
losses. Of course these are rather theoretical considerations: In
practice there are restrictions to the size of the state (in terms of
photon number) which can be experimentally generated. So even if the
curves drops to $0.5$ for $N\rightarrow\infty$, interferometry using
``smaller'' quantum states has a significant advantage both in terms
of scaling and absolute precision over classical interferometry.
Particularly for very small numbers of photons (unbalanced) N00N
states are optimal.  Above this threshold the use of the more complex,
optimized states or one of our experimentally more feasible
alternatives is favourable.

This research was supported by the EPSRC (UK) through the QIP IRC
(GR/S82176/01), the AFOSR through the EOARD, the European Commission
under the Integrated Project QAP (Contract No.~015848), the Royal
Society and the Polish MNISW (N N202 1489 33).


\begin{thebibliography}{18}
\expandafter\ifx\csname natexlab\endcsname\relax\def\natexlab#1{#1}\fi
\expandafter\ifx\csname bibnamefont\endcsname\relax
  \def\bibnamefont#1{#1}\fi
\expandafter\ifx\csname bibfnamefont\endcsname\relax
  \def\bibfnamefont#1{#1}\fi
\expandafter\ifx\csname citenamefont\endcsname\relax
  \def\citenamefont#1{#1}\fi
\expandafter\ifx\csname url\endcsname\relax
  \def\url#1{\texttt{#1}}\fi
\expandafter\ifx\csname urlprefix\endcsname\relax\def\urlprefix{URL }\fi
\providecommand{\bibinfo}[2]{#2}
\providecommand{\eprint}[2][]{\url{#2}}

\bibitem[{\citenamefont{Giovannetti et~al.}(2004)\citenamefont{Giovannetti,
  Lloyd, and Maccone}}]{Giovanneti04}
\bibinfo{author}{\bibfnamefont{V.}~\bibnamefont{Giovannetti}},
  \bibinfo{author}{\bibfnamefont{S.}~\bibnamefont{Lloyd}}, \bibnamefont{and}
  \bibinfo{author}{\bibfnamefont{L.}~\bibnamefont{Maccone}},
  \bibinfo{journal}{Science} \textbf{\bibinfo{volume}{306}},
  \bibinfo{pages}{1330} (\bibinfo{year}{2004}).

\bibitem[{\citenamefont{Giovannetti et~al.}(2006)\citenamefont{Giovannetti,
  Lloyd, and Maccone}}]{Giovanneti06}
\bibinfo{author}{\bibfnamefont{V.}~\bibnamefont{Giovannetti}},
  \bibinfo{author}{\bibfnamefont{S.}~\bibnamefont{Lloyd}}, \bibnamefont{and}
  \bibinfo{author}{\bibfnamefont{L.}~\bibnamefont{Maccone}},
  \bibinfo{journal}{Phys. Rev. Lett.} \textbf{\bibinfo{volume}{96}},
  \bibinfo{pages}{010401} (\bibinfo{year}{2006}).

\bibitem[{\citenamefont{Higgins et~al.}(2007)\citenamefont{Higgins, Berry,
  Bartlett, Wiseman, and Pryde}}]{Higgins07}
\bibinfo{author}{\bibfnamefont{B.~L.} \bibnamefont{Higgins}},
  \bibinfo{author}{\bibfnamefont{D.~W.} \bibnamefont{Berry}},
  \bibinfo{author}{\bibfnamefont{S.~D.} \bibnamefont{Bartlett}},
  \bibinfo{author}{\bibfnamefont{H.~M.} \bibnamefont{Wiseman}},
  \bibnamefont{and} \bibinfo{author}{\bibfnamefont{G.~J.} \bibnamefont{Pryde}},
  \bibinfo{journal}{Nature} \textbf{\bibinfo{volume}{450}},
  \bibinfo{pages}{393} (\bibinfo{year}{2007}).

\bibitem[{\citenamefont{Mitchell et~al.}(2004)\citenamefont{Mitchell, Lundeen,
  and Steinberg}}]{Mitchell04}
\bibinfo{author}{\bibfnamefont{M.~W.} \bibnamefont{Mitchell}},
  \bibinfo{author}{\bibfnamefont{J.~S.} \bibnamefont{Lundeen}},
  \bibnamefont{and} \bibinfo{author}{\bibfnamefont{J.~S.}
  \bibnamefont{Steinberg}}, \bibinfo{journal}{Nature}
  \textbf{\bibinfo{volume}{429}}, \bibinfo{pages}{161} (\bibinfo{year}{2004}).

\bibitem[{\citenamefont{Eisenberg et~al.}(2005)\citenamefont{Eisenberg,
  Hodelin, Khoury, and Bouwmeester}}]{Eisenberg05}
\bibinfo{author}{\bibfnamefont{H.~S.} \bibnamefont{Eisenberg}},
  \bibinfo{author}{\bibfnamefont{J.~F.} \bibnamefont{Hodelin}},
  \bibinfo{author}{\bibfnamefont{G.}~\bibnamefont{Khoury}}, \bibnamefont{and}
  \bibinfo{author}{\bibfnamefont{D.}~\bibnamefont{Bouwmeester}},
  \bibinfo{journal}{Phys. Rev. Lett.} \textbf{\bibinfo{volume}{94}},
  \bibinfo{pages}{090502} (\bibinfo{year}{2005}).

\bibitem[{\citenamefont{Nagata et~al.}(2007)\citenamefont{Nagata, Okamoto,
  O'Brien, Sasaki, and Takeuchi}}]{Nagata07}
\bibinfo{author}{\bibfnamefont{T.}~\bibnamefont{Nagata}},
  \bibinfo{author}{\bibfnamefont{R.}~\bibnamefont{Okamoto}},
  \bibinfo{author}{\bibfnamefont{J.}~\bibnamefont{O'Brien}},
  \bibinfo{author}{\bibfnamefont{K.}~\bibnamefont{Sasaki}}, \bibnamefont{and}
  \bibinfo{author}{\bibfnamefont{S.}~\bibnamefont{Takeuchi}},
  \bibinfo{journal}{Science} \textbf{\bibinfo{volume}{316}},
  \bibinfo{pages}{726} (\bibinfo{year}{2007}).

\bibitem[{\citenamefont{Resch et~al.}(2007)}]{Resch07}
\bibinfo{author}{\bibfnamefont{K.~J.} \bibnamefont{Resch}}
  \bibnamefont{et~al.}, \bibinfo{journal}{Phys. Rev. Lett.}
  \textbf{\bibinfo{volume}{98}}, \bibinfo{pages}{223601}
  (\bibinfo{year}{2007}).

\bibitem[{\citenamefont{Huelga et~al.}(1997)}]{Huelga97}
\bibinfo{author}{\bibfnamefont{S.~F.} \bibnamefont{Huelga}}
  \bibnamefont{et~al.}, \bibinfo{journal}{Phys. Rev. Lett.}
  \textbf{\bibinfo{volume}{79}}, \bibinfo{pages}{3865} (\bibinfo{year}{1997}).

\bibitem[{\citenamefont{Shaji and Caves}(2007)}]{Shaji07}
\bibinfo{author}{\bibfnamefont{A.}~\bibnamefont{Shaji}} \bibnamefont{and}
  \bibinfo{author}{\bibfnamefont{C.~M.} \bibnamefont{Caves}},
  \bibinfo{journal}{Phys. Rev. A} \textbf{\bibinfo{volume}{76}},
  \bibinfo{pages}{032111} (\bibinfo{year}{2007}).

\bibitem[{\citenamefont{Sahovar and Milburn}(2006)}]{Sarovar06}
\bibinfo{author}{\bibfnamefont{M.}~\bibnamefont{Sahovar}} \bibnamefont{and}
  \bibinfo{author}{\bibfnamefont{G.~J.} \bibnamefont{Milburn}},
  \bibinfo{journal}{J. Phys. A} \textbf{\bibinfo{volume}{39}},
  \bibinfo{pages}{8487} (\bibinfo{year}{2006}).

\bibitem[{\citenamefont{Gilbert et~al.}(2006)\citenamefont{Gilbert, Hamrick and 
  Weinstein}}]{Gilbert06}
\bibinfo{author}{\bibfnamefont{G.}~\bibnamefont{Gilbert}},
  \bibinfo{author}{\bibfnamefont{M.}~\bibnamefont{Hamrick}},
  \bibnamefont{and} \bibinfo{author}{\bibfnamefont{Y.~S.}
  \bibnamefont{Weinstein}} (\bibinfo{year}{2006}), \eprint{arXiv:quant-ph/0612156v1}.

\bibitem[{\citenamefont{Huver et~al.}(2008)\citenamefont{Huver, Wildfeuer, and
  Dowling}}]{Huver08}
\bibinfo{author}{\bibfnamefont{S.~D.} \bibnamefont{Huver}},
  \bibinfo{author}{\bibfnamefont{C.~F.} \bibnamefont{Wildfeuer}},
  \bibnamefont{and} \bibinfo{author}{\bibfnamefont{J.~P.}
  \bibnamefont{Dowling}} (\bibinfo{year}{2008}), \eprint{arXiv:0805.0296v1 [quant-ph]}.

\bibitem[{\citenamefont{Bollinger et~al.}(1996)\citenamefont{Bollinger, Itano,
  Wineland, and Heinzen}}]{Bollinger96}
\bibinfo{author}{\bibfnamefont{J.~J.} \bibnamefont{Bollinger}},
  \bibinfo{author}{\bibfnamefont{W.~M.} \bibnamefont{Itano}},
  \bibinfo{author}{\bibfnamefont{D.~J.} \bibnamefont{Wineland}},
  \bibnamefont{and} \bibinfo{author}{\bibfnamefont{D.~J.}
  \bibnamefont{Heinzen}}, \bibinfo{journal}{Phys. Rev. A}
  \textbf{\bibinfo{volume}{54}}, \bibinfo{pages}{R4649} (\bibinfo{year}{1996}).

\bibitem[{\citenamefont{Braunstein and Caves}(1994)}]{Braunstein94}
\bibinfo{author}{\bibfnamefont{S.~L.} \bibnamefont{Braunstein}}
  \bibnamefont{and} \bibinfo{author}{\bibfnamefont{C.~M.} \bibnamefont{Caves}},
  \bibinfo{journal}{Phys. Rev. Lett.} \textbf{\bibinfo{volume}{72}},
  \bibinfo{pages}{3439} (\bibinfo{year}{1994}); 
\bibinfo{author}{\bibfnamefont{S.~L.} \bibnamefont{Braunstein}},
  \bibinfo{author}{\bibfnamefont{C.~M.} \bibnamefont{Caves}}, \bibnamefont{and}
  \bibinfo{author}{\bibfnamefont{G.~J.} \bibnamefont{Milburn}},
  \bibinfo{journal}{Ann. Phys. (NY)} \textbf{\bibinfo{volume}{247}},
  \bibinfo{pages}{135} (\bibinfo{year}{1996}).

\bibitem[{\citenamefont{Helstrom}(1976)}]{Helstrom}
\bibinfo{author}{\bibfnamefont{C.~W.} \bibnamefont{Helstrom}},
  \emph{\bibinfo{title}{Quantum Detection and Estimation Theory}}
  (\bibinfo{publisher}{Academic}, \bibinfo{address}{New York},
  \bibinfo{year}{1976}).

\bibitem[{\citenamefont{Holevo}(1982)}]{Holevo}
\bibinfo{author}{\bibfnamefont{A.~S.} \bibnamefont{Holevo}},
  \emph{\bibinfo{title}{Probabilistic and Statistical Aspects of Quantum
  Theory}} (\bibinfo{publisher}{North-Holland}, \bibinfo{address}{Amsterdam},
  \bibinfo{year}{1982}).

\bibitem[{\citenamefont{Demkowicz-Dobrzanski et~al.}()}]{Demkowicz08}
\bibinfo{author}{\bibfnamefont{R.}~\bibnamefont{Demkowicz-Dobrzanski}}
  \bibnamefont{et~al.}, \bibinfo{note}{in preparation}.

\bibitem[{\citenamefont{M\o{}lmer}(1997)}]{Molmer97}
\bibinfo{author}{\bibfnamefont{K.}~\bibnamefont{M\o{}lmer}},
  \bibinfo{journal}{Phys. Rev. A} \textbf{\bibinfo{volume}{55}},
  \bibinfo{pages}{3195} (\bibinfo{year}{1997}).

\bibitem[{\citenamefont{Fujiwara}(2001)}]{Fujiwara01}
\bibinfo{author}{\bibfnamefont{A.}~\bibnamefont{Fujiwara}},
  \bibinfo{journal}{Phys. Rev. A} \textbf{\bibinfo{volume}{63}},
  \bibinfo{pages}{042304} (\bibinfo{year}{2001}).

\bibitem[{\citenamefont{Holland and Burnett}(1993)}]{Holland93}
\bibinfo{author}{\bibfnamefont{M.~J.} \bibnamefont{Holland}} \bibnamefont{and}
  \bibinfo{author}{\bibfnamefont{K.}~\bibnamefont{Burnett}},
  \bibinfo{journal}{Phys. Rev. Lett.} \textbf{\bibinfo{volume}{71}},
  \bibinfo{pages}{1355} (\bibinfo{year}{1993});  
\bibinfo{author}{\bibfnamefont{J.~A.} \bibnamefont{Dunningham}},
  \bibinfo{author}{\bibfnamefont{K.}~\bibnamefont{Burnett}}, \bibnamefont{and}
  \bibinfo{author}{\bibfnamefont{S.~M.} \bibnamefont{Barnett}},
  \bibinfo{journal}{Phys. Rev. Lett.} \textbf{\bibinfo{volume}{89}},
  \bibinfo{pages}{150401} (\bibinfo{year}{2002}).

\end{thebibliography}
\end{document}